\documentstyle[pre,aps]{revtex}

\tighten

\begin{document}

\draft
\title{
Charge superselection rule does not rule out 
pure states of subsystems
to be coherent superpositions of states with different charges
}
\author{
Akira Shimizu\cite{shmz} and Takayuki Miyadera
}
\address{
Department of Basic Science, University of Tokyo, 
Komaba, Meguro-ku, Tokyo 153-8902, Japan
}
\date{Received 23 February 2001}
\maketitle
\begin{abstract}
We consider a huge quantum system that is subject to the charge superselection 
rule, which requires that any pure state must be 
an eigenstate of the total charge.
We regard some parts of the system as ``subsystems" 
S$_1$, S$_2$, $\cdots$, S$_M$,
and the rest as an environment E.
We assume that one does not measure anything of E, i.e., 
one is only interested in observables of 
the joint subsystem S $\equiv {\rm S}_1 + {\rm S}_2 + \cdots + {\rm S}_M$.
We show that there exist states 
$| \Phi \rangle_{\rm tot}$
with the following properties:
(i) The reduced density operator
$
{\rm Tr}_{\rm E}
\left( | \Phi \rangle_{\rm tot} \ \null_{\rm tot} \langle \Phi | \right)
$ is completely equivalent to a vector state
$| \varphi \rangle_{\rm S} \ \null_{\rm S} \langle \varphi |$ 
of S, 
for any gauge-invariant observable of S.
(ii) $| \varphi \rangle_{\rm S}$
is a simple product of vector states of individual subsystems;
$
| \varphi \rangle_{\rm S}
=
| C^{(1)} \rangle_1 | C^{(2)} \rangle_2 \cdots
$, 
where
$
| C^{(k)} \rangle_k 
$
is a vector state in S$_k$ which is {\it not} an
eigenstate of the charge in S$_k$.
Furthermore, 
one can associate to each subsystem S$_k$ 
the vector state $| C^{(k)} \rangle_k$
and observables which are {\em not} necessarily 
gauge invariant in each subsystem,  
and $| C^{(k)} \rangle_k$ is then a pure state.
These results justify taking (a) 
superpositions of states with different charges, and 
(b) non-gauge-invariant operators, such as 
the order parameter of the breaking of the gauge symmetry, as observables,
for subsystems. 
\end{abstract}


In quantum theory, 
some superpositions of states are not permitted as pure states \cite{haag}.
In particular, the charge superselection rule (CSSR)
forbids coherent superpositions of states with different charges \cite{haag}.
Namely, any pure state must be an eigenstate of the total charge
$\hat N_{\rm tot}$.
However, it is customary to take such superpositions when 
one discusses the breaking of a gauge symmetry.
Superconductors and superfluids are typical examples.
If the system size $V$ is infinite, 
this does not conflict with the CSSR
because $\hat N_{\rm tot}$ becomes ill-defined as $V \to \infty$, 
and the CSSR becomes inapplicable.
In real physical systems, however, 
phase transitions practically occur 
for finite ($V < + \infty$) systems as well.
In particular, 
phase transitions have been 
observed in relatively small systems, 
including 
small superconductors \cite{super}, 
Helium atoms in a micro bubble \cite{He4bubble}, and 
laser-trapped atoms \cite{atom}.
The meaning of the symmetry breaking in such systems
has been a subject of active research \cite{andrews,java,barnett,SMprl2000}.
The purpose of this paper is to present a general discussion which justifies 
taking coherent superpositions of states with different charges
for finite quantum systems subject to the CSSR, such as 
charged particles and massive bosons.
Furthermore, we also justify 
taking 
non-gauge-invariant operators such as 
the order parameter of the breaking of the gauge symmetry, as observables
of subsystems.


We consider a huge quantum system that is subject to the CSSR.
We regard some parts of the system as ``subsystems" S$_1$, S$_2$, $\cdots$, S$_M$,
and the rest as the environment E.
We assume the usual situation where 
(i) E is much larger than the joint subsystem 
S $\equiv {\rm S}_1 + {\rm S}_2 + \cdots + {\rm S}_M$,
and 
(ii) one is not interested in (thus, one will not measure) 
degrees of freedom of E, i.e, one is only interested in S
(or some parts of S).
The Hilbert space ${\cal H}_{\rm tot}$ of the total system 
is the tensor product
of the Hilbert spaces of S$_1$, S$_2$, $\cdots$, S$_M$ and E;
$ 
{\cal H}_{\rm tot}
=
{\cal H}_{\rm S}
\otimes 
{\cal H}_{\rm E},
$ 
where
$ 
{\cal H}_{\rm S}
\equiv
{\cal H}_1 \otimes {\cal H}_2 \otimes \cdots \otimes {\cal H}_M.
$ 
The total charge (in some unit) $\hat N_{\rm tot}$ is the
sum of the charges of S$_1$, S$_2$, $\cdots$, S$_M$ and E;
\begin{equation}
\hat N_{\rm tot}
=
\sum_{k=1}^M \hat N_k + \hat N_{\rm E}.
\end{equation}
Products of eigenfunctions
$| N_1 n_1 \rangle_1$, 
$\cdots$, $| N_M n_M \rangle_M$, and 
$| N_{\rm E} \ell \rangle_{\rm E}$
of
$\hat N_1$, 
$\cdots$, $\hat N_M$, and $\hat N_{\rm E}$,
respectively, 
form complete basis sets of ${\cal H}_{\rm tot}$.
Here, 
$N_k$ ($k = 1, 2, \cdots, M$) and $N_{\rm E}$ are eigenvalues of 
$\hat N_k$ and $\hat N_{\rm E}$,
respectively, 
and 
$n_k$ and $\ell$ denote additional quantum numbers.

The CSSR requires that 
any pure state of the total system must be an 
eigenstate of $\hat N_{\rm tot}$, i.e., 
superposition is allowed only among states with a fixed value of 
the eigenvalue $N_{\rm tot}$ of $\hat N_{\rm tot}$.
Consider the following state that satisfies
this requirement:
\begin{equation}
| \Phi \rangle_{\rm tot}
=
\sum_{N_1,n_1} \cdots \sum_{N_M,n_M}
\sum_{\ell}
C^{(1)}_{N_1 n_1} \cdots C^{(M)}_{N_M n_M}
C^{({\rm E})}_{N_{\rm S} \ell} \
| N_1 n_1 \rangle_1 \cdots | N_M n_M \rangle_M
| N_{\rm tot}- N_{\rm S}, \ell \rangle_{\rm E},
\label{Phi}\end{equation}
where
$N_{\rm S} = \sum_k N_k$, and the superposition 
coefficients are normalized as
\begin{equation}
\sum_{N,n} |C^{(k)}_{N n}|^2 
=
\sum_{\ell} |C^{({\rm E})}_{N_{\rm S} \ell}|^2
= 1.
\label{norm_C}\end{equation}
For this state, the probability of finding 
$N_{\rm tot}-N_{\rm S}$ bosons in E takes
almost the same values for all $N_{\rm S}$ such that 
$ 
|N_{\rm S} - \langle N_{\rm S} \rangle |
<
\langle \delta N_{\rm S}^2 \rangle^{1/2}
$ \cite{range}. 
This property seems natural for a huge environment.
Since we assume that one will not measure degrees of freedom of E, 
we are interested in 
the reduced density operator $\hat \rho_{\rm S}$ of S, 
which is evaluated as
\begin{eqnarray}
&&
\hat \rho_{\rm S} = {\rm Tr}_{\rm E}
\left( | \Phi \rangle_{\rm tot} \ \null_{\rm tot} \langle \Phi | \right)
\nonumber\\
&& \quad =
\sum_{N'_1,n'_1} \cdots \sum_{N'_M,n'_M} 
\sum_{N_1,n_1} \cdots \sum_{N_M,n_M}
\delta_{N_1+\cdots+N_M, \ N'_1+\cdots+N'_M}
\
C^{(1)}_{N'_1 n'_1} \cdots C^{(M)}_{N'_M n'_M} 
C^{(M)*}_{N_M n_M} \cdots C^{(1)*}_{N_1 n_1}
\nonumber\\
&& \qquad 
\times \
| N'_1 n'_1 \rangle_1 \cdots | N'_M n'_M \rangle_M
\null_M \langle N_M n_M | \cdots \null_1 \langle N_1 n_1 |.
\label{rho}\end{eqnarray}
We can easily show that
$ 
(\hat \rho_{\rm S})^2 \neq \hat \rho_{\rm S}
$ 
unless
\begin{equation}
\sum_{n}|C^{(k)}_{Nn}|^2 = \delta_{N,N^{(k)}_0}
\mbox{ for all $k$}
\end{equation}
for some set of numbers $N^{(1)}_0, \cdots, N^{(M)}_0$.
We exclude this trivial case (where the CSSR is satisfied
in each subsystem) from our consideration.
Then, $(\hat \rho_{\rm S})^2 \neq \hat \rho_{\rm S}$, 
and one may say that
$\hat \rho_{\rm S}$ represents a mixed state.
However, the relation $(\hat \rho_{\rm S})^2 \neq \hat \rho_{\rm S}$
only ensures that 
for any vector state $| \varphi \rangle_{\rm S}$ 
($\in {\cal H}_{\rm S}$)
there exists
some {\em operator} $\hat \Xi_{\rm S}$ 
(on ${\cal H}_{\rm S}$)
for which 
\begin{equation}
{\rm Tr}_{\rm S} \left( \hat \rho_{\rm S} \hat \Xi_{\rm S} \right)
\neq
\null_{\rm S} \langle \varphi |
\hat \Xi_{\rm S}
| \varphi \rangle_{\rm S}.
\end{equation}
Note that such a general operator $\hat \Xi_{\rm S}$ 
is not necessarily gauge-invariant.
Hence, $\hat \Xi_{\rm S}$ might not be an {\em observable}, 
which must be gauge-invariant.
%
In fact, we first show that
$\hat \rho_{\rm S}$ is equivalent to 
a vector state
$| \varphi \rangle_{\rm S}$
for all gauge-invariant (thus physical) observables on ${\cal H}_{\rm S}$.
This statement might not sound surprising because
a vector state is not necessarily a pure state \cite{haag,vs}.
[Here, we use the precise definition of pure and mixed states \cite{precise}, 
rather than misleading definitions such as 
$\hat \rho^2 = \hat \rho$ and $\hat \rho^2 \neq \hat \rho$.] 
In fact, the equivalence of
$| \varphi \rangle_{\rm S}$
to 
$\hat \rho_{\rm S}$ 
means that 
the vector state $| \varphi \rangle_{\rm S}$ is 
a mixed state.
In other words, 
$| \varphi \rangle_{\rm S}$ is a vector state in a 
{\em reducible} representation of the algebra of 
gauge-invariant observables \cite{vs}.

Actually, we first show a stronger statement:
$| \varphi \rangle_{\rm S}$
is a simple product of vector states of individual subsystems;
\begin{equation}
\hat \rho_{\rm S}
\mbox{ is equivalent to }
| \varphi \rangle_{\rm S}
=
| C^{(1)} \rangle_1 | C^{(2)} \rangle_2 \cdots | C^{(M)} \rangle_M
\mbox{ for any gauge-invariant observables in 
${\cal H}_{\rm S}$},
\label{equivalance}\end{equation}
where $| C^{(k)} \rangle_k$ is a coherent superposition of 
states with different charges;
\begin{equation}
| C^{(k)} \rangle_k 
=
\sum_{N,n}
C^{(k)}_{Nn}
| N n \rangle_k.
\end{equation}
To see this, we recall that 
one will not measure degrees of freedom of E.
This means that 
one measures only observables which take the following form;
\begin{equation}
\hat A_{\rm S} \otimes \hat 1_{\rm E},
\end{equation}
where 
$\hat A_{\rm S}$ is an operator on ${\cal H}_{\rm S}$, 
and $\hat 1_{\rm E}$ denotes the unity operator on 
${\cal H}_{\rm E}$.
Note that
$\hat A_{\rm S} \otimes \hat 1_{\rm E}$
must be gauge-invariant
 because of the gauge invariance of the total system, 
hence
$\hat A_{\rm S}$ must also be gauge-invariant.
This requires that 
$N_{\rm S}$ ($= \sum_k N_k$) should be conserved
by the operation of $\hat A_{\rm S}$. 
Hence, the matrix elements of $\hat A_{\rm S}$ should take 
the following form;
\begin{equation}
\null_1 \langle N_1 n_1 | \cdots \null_M \langle N_M n_M |
\ \hat A_{\rm S} \
| N'_M n'_M \rangle_M \cdots | N'_1 n'_1 \rangle_1
=
\delta_{N_1+\cdots+N_M, N'_1+\cdots+N'_M}
A^{N_1 n_1 \cdots N_M n_M}_{N'_1 n'_1 \cdots N'_M n'_M}.
\label{A}\end{equation}
Hence, the expectation value of $\hat A_{\rm S}$ for 
$\hat \rho_{\rm S}$ is evaluated as
\begin{eqnarray}
&&
\langle A_{\rm S} \rangle
=
{\rm Tr}_{\rm S} \left( \hat \rho_{\rm S} \hat A_{\rm S} \right)
\nonumber\\
&& \quad =
\sum_{N'_1,n'_1} \cdots \sum_{N'_M,n'_M} 
\sum_{N_1,n_1} \cdots \sum_{N_M,n_M}
C^{(1)}_{N'_1 n'_1} \cdots C^{(M)}_{N'_M n'_M}
C^{(M)*}_{N_M n_M} \cdots C^{(1)*}_{N_1 n_1}
\delta_{N_1+\cdots+N_M, N'_1+\cdots+N'_M}
A^{N_1 n_1 \cdots N_M n_M}_{N'_1 n'_1 \cdots N'_M n'_M}
\nonumber\\
&& \quad =
\null_{\rm S} \langle \varphi |
\hat A_{\rm S}
| \varphi \rangle_{\rm S},
\end{eqnarray}
and Eq.\ (\ref{equivalance}) is proved.
The point is that Eqs.\ (\ref{rho}) and (\ref{A}) contain the same
Kronecker's delta, 
$\delta_{N_1+\cdots+N_M, N'_1+\cdots+N'_M}$.
Although this factor in Eq.\ (\ref{rho}) makes 
$\hat \rho_{\rm S}$ different from 
$
| \varphi \rangle_{\rm S} \ \null_{\rm S}\langle \varphi |
$, 
the difference becomes totally irrelevant to 
$
\langle A_{\rm S} \rangle
$
because of the same factor in Eq.\ (\ref{A}).
(Recall that 
$
(\delta_{N_1+\cdots+N_M, N'_1+\cdots+N'_M})^2
=
\delta_{N_1+\cdots+N_M, N'_1+\cdots+N'_M}
$.)

Moreover, we can easily show that 
$\hat \rho_{\rm S-S_M} \equiv {\rm Tr}_M [\hat \rho_{\rm S}]$ 
is also equivalent to a vector state,
$
| C^{(1)} \rangle_1 | C^{(2)} \rangle_2 \cdots  | C^{(M)} \rangle_{M-1}
$,
where 
${\rm Tr}_M$ denotes the trace operation over ${\cal H}_M$.
This result is easily generalized:
For states of the form of (\ref{Phi}), 
its reduced density operator for any set of subsystems
is completely equivalent to a vector state, which is a simple 
product of vector states of the individual subsystems, 
if one measures only gauge-invariant observables of the subsystems.
%
Note that there is no `entanglement' between any subsystems \cite{ent}.
%
In general, the 
absence of entanglement among subsystems means the `separability',
i.e., it is possible to 
control quantum states of individual subsystems independently 
by local operations \cite{peres}.
Hence, 
one can control 
the state 
$
| C^{(k)} \rangle_k
$ of each subsystem by local operations,
without perturbing the other subsystems \cite{perturbed}.


For example,
for interacting many bosons, 
which are confined in a large but {\em finite} box,
one may take 
$
| C^{(k)} \rangle_k
$
to be the `coherent state of interacting bosons' (CSIB), 
which is defined by \cite{SMprl2000,SI,complicated}
\begin{equation}
| \alpha_k, {\rm G} \rangle_k
=
e^{- |\alpha_k|^2/2}
\sum_{N=0}^{\infty}
\frac{\alpha_k^N}{\sqrt{N!}}
|N, {\rm G} \rangle_k,
\label{CSIBk}\end{equation}
where $\alpha_k = |\alpha_k| e^{i \phi_k}$ is a complex amplitude, and
$|N, {\rm G} \rangle_k$ denotes 
the ground state that has {\em exactly} $N$ bosons, 
which we call the `number state of interacting bosons' (NSIB)
\cite{SMprl2000,SI}.
One may also take 
$
| C^{(l)} \rangle_l
$ 
as $| \alpha_l, {\rm G} \rangle_l$, where $\alpha_l = |\alpha_l| e^{i \phi_l}$. Then, $\hat \rho_{\rm S}$ is equivalent to
\begin{equation}
| \varphi \rangle_{\rm S}
=
\cdots
| \alpha_k, {\rm G} \rangle_k
\cdots
| \alpha_l, {\rm G} \rangle_l
\cdots
\label{CSIBkl}\end{equation}
This state has an almost definite value of the 
relative phase $\phi_{kl}$ between two condensates in S$_k$ and S$_l$.
To see this, 
we may define the operator
(acting on ${\cal H}_k \otimes {\cal H}_l$) corresponding to $e^{i \phi_{kl}}$ 
by
\begin{equation}
\widehat{e^{i \phi_{kl}}}
\equiv
\widehat{e^{i \phi_k}}
\
(\widehat{e^{i \phi_l}})^\dagger.
\label{ephikl}\end{equation}
Here, $\widehat{e^{i \phi_k}}$, acting on ${\cal H}_k$, 
is not an exponential of 
some phase operator $\hat \phi_k$, but is defined by
\begin{equation}
\widehat{e^{i \phi_k}}
\equiv
(\hat b_{0k}^\dagger \hat b_{0k} + 1)^{-1/2} b_{0k},
\end{equation}
where $\hat b_{0k}$ denotes the operator $\hat b_0$, which is defined 
in Ref.\ \cite{SI} as a nonlinear function of free operators, for S$_k$.
This operator is a linear combination of the
cosine and sine operators of Ref.\ \cite{SI}.
In the same way, $\widehat{e^{i \phi_l}}$, acting on ${\cal H}_l$,
is defined using 
$\hat b_{0l}$, which is $\hat b_0$ for S$_l$.
By similar calculations as in Ref.\ \cite{SI}, 
we can show that 
$\widehat{e^{i \phi_k}}$, $(\widehat{e^{i \phi_l}})^\dagger$, and
$\widehat{e^{i \phi_{kl}}}$
have almost definite values, and 
we can regard $\phi_{kl} \simeq \phi_k - \phi_l$.
Hence, the state (\ref{CSIBkl}) has the almost definite value of the 
relative phase.
It should not be confused with states of the following form, 
which have been frequently discussed in the literature \cite{ex-ent};
\begin{equation}
|\varphi_{\rm ent} \rangle_{\rm S}
=
\sum_N C_N \sum_{N_k}
\cdots e^{i N_k \phi_k} R_{N_k}  |N_k, {\rm G} \rangle_k 
\cdots 
e^{i (N-N_k) \phi_l} R_{N-N_k} |N-N_k, {\rm G} \rangle_l \cdots,
\label{ent}\end{equation}
where $C_N$ is a complex coefficient and $R$'s are real ones.
This state also has an almost definite relative phase
if $R$'s are appropriately taken.
However, since S$_k$ and S$_l$ are strongly 
entangled in this state \cite{ent}, 
one cannot control the state of S$_k$ by local operations
on S$_k$, without perturbing the state of S$_l$.
In contrast, this is possible for 
the simple product state $|\varphi \rangle_{\rm S}$.
Moreover, we now show that $|\varphi \rangle_{\rm S}$ 
allows us to treat observables of individual subsystems separately.


As exemplified by $\widehat{e^{i \phi_{kl}}}$ above, 
$\hat A_{\rm S}$ is generally 
a sum of products of operators of subsystems \cite{hc};
\begin{equation}
\hat A_{\rm S}
=
\sum
\hat A_{k} \hat A_{l} \cdots,
\end{equation}
where
$\hat A_{k}$ ($\hat A_{l}$) is an operator on ${\cal H}_k$ (${\cal H}_l$), 
and the sum is over combinations of 
$\hat A_{k} \hat A_{l} \cdots$.
For the product state $|\varphi \rangle_{\rm S}$, 
the expectation value of $\hat A_{\rm S}$ 
is simply evaluated from 
the expectation values of individual subsystems as
\begin{equation}
{}_{\rm S} \langle \varphi | \hat A_{\rm S} |\varphi \rangle_{\rm S}
=
\sum
{}_k \langle C^{(k)} | \hat A_k |C^{(k)} \rangle_k
\
{}_l \langle C^{(l)} | \hat A_l |C^{(l)} \rangle_l
\cdots.
\label{exp-AS}\end{equation}
From Eqs.\ (\ref{equivalance}) and (\ref{exp-AS}), 
{\em we can consider each subsystem separately} from 
the other subsystems.
Namely, for subsystem S$_k$, 
we can consider that its state is the vector state $| C^{(k)} \rangle_k$, 
and that $\hat A_{k}$ is one of its observables.
When the expectation value of $\hat A_{\rm S}$ 
for the set of subsystems is necessary, it can be evaluated 
from the expectation values of individual subsystems, as
Eq.\ (\ref{exp-AS}).
Note that the gauge invariance of $\hat A_{\rm S}$ 
does {\it not} require the gauge invariance of
$\hat A_{k}$ of each subsystem:
It rather requires that 
only gauge-invariant {\em combinations} of $\hat A_{k} \hat A_{l} \cdots$
should appear in the sum.
[Namely, each $N_k$ can vary by the operation of $\hat A_{k}$, 
whereas 
$N_{\rm S}$ ($= \sum_k N_k$) 
is conserved by the operation of $\hat A_{k} \hat A_{l} \cdots$.]
For example, 
$\widehat{e^{i \phi_{kl}}}$ is gauge invariant, whereas 
neither $\widehat{e^{i \phi_{k}}}$ nor $\widehat{e^{i \phi_{l}}}$ is.
Hence, by considering $\hat A_{k}$ as an observable of S$_k$, 
we can include 
non-gauge invariant operators into the set of 
observables in ${\cal H}_k$, 
with the restriction that 
among various combinations of
$\hat A_{k} \hat A_{l} \cdots$
we must take only gauge-invariant ones as physical combinations:
Results for other combinations should be discarded.
This formulation gives correct results for all 
gauge-invariant (thus physical) combinations.
The vector state $| C^{(k)} \rangle_k$ then becomes a pure state
of an irreducible representation of the algebra of
observables of S$_k$, because non-gauge invariant observables are 
now included in the algebra \cite{vs}.
We have thus arrived at {\em non-gauge invariant observables
and a pure state which is a superposition of states with different charges, 
for each subsystem}.
%



The restriction that 
we must take only gauge-invariant ones 
among various combinations of
$\hat A_{k} \hat A_{l} \cdots$ has the following physical meaning:
If $\hat A_{k}$ is a non-gauge invariant observable of S$_k$, then 
its value can only be defined relative to some reference 
observable $\hat A_{l}$ of some reference system S$_l$.
When $| \varphi \rangle_{\rm S}$ takes the form 
of Eq.\ (\ref{CSIBkl}), for example, 
we can consider that S$_k$ is in a pure state 
$| \alpha_k, {\rm G} \rangle_k$, and
that the phase factor 
$\widehat{e^{i \phi_{k}}}$ is one of observables on ${\cal H}_k$.
However, 
like the classical phase factor, 
the quantum phase factor $\widehat{e^{i \phi_{k}}}$ can 
only be defined relative to some reference
(such as $(\widehat{e^{i \phi_{l}}})^\dagger$ of S$_l$).
In contrast to the position observable, 
which also requires a reference system but it can be any system composed of 
any material, 
the reference system of the phase factor should 
contain particles of the same kind as the particles in S$_k$.
Hence, the system of interest (S$_k$) and the reference system (S$_l$)
should be subsystems of a larger system of the same kind of 
particles.
When the larger system contains a huge system(s), 
of which we are not interested in, 
we may call it an environment E, 
and the present model is applicable.
When one is only interested in S$_k$, 
the reference system S$_l$ may be considered as 
a part of the external systems, which include an apparatus 
with which one measures $\phi_{kl}$.
In this case, 
the reference system S$_l$ can be regarded as a part of 
the measuring apparatus.
(In the case of measurement of the position, 
for example, a material which defines the origin of the
position coordinate can be considered as a part of the measuring apparatus
of the position.)
Although $\widehat{e^{i \phi_{k}}}$ is not gauge invariant, 
results of measurement is gauge invariant because
$\widehat{e^{i \phi_k}} \ (\widehat{e^{i \phi_l}})^\dagger$ 
is gauge invariant.
In general terms, 
although results of any physical {\it measurements} must be gauge invariant,
it does not necessarily mean that 
any {\it observables} of a subsystem must be gauge-invariant, 
because the gauge invariance of $\hat A_{k} \hat A_{l}$
(which ensures the gauge invariance of results of measurements)
does not necessarily mean the gauge invariance of $\hat A_{k}$.


Finally, we present some significant applications of the present results.
First, Our results give a natural interpretation of 
the order parameter $\hat O$ of a finite system, 
which can exhibit 
the breaking of the gauge symmetry in the infinite-volume limit.
Although one might suspect that $\hat O$ must be excluded from observables 
since it is not gauge invariant, 
our results allow one to include it among observables as one of $\hat A_{k}$'s.
Second, one might also suspect that the CSSR would 
forbid a pure state which has a finite expectation value of
$\hat O$ because its expectation value vanishes for any eigenstate 
of $\hat N$.
However, our results show that such a pure state is allowed as
a state of a subsystem, 
$
| C^{(k)} \rangle_k
$, 
which is not entangled with states of the other subsystems
The absence of entanglement allows the preparation of
such a pure state of a subsystem without perturbing 
the other subsystems (with perturbing the environment only).
For interacting many bosons, for example, 
the order parameter of the condensation is usually taken as
the field operator $\hat \psi$ of bosons.
The present work justifies taking 
such a non-gauge-invariant operator as an observable of a subsystem.
The expectation value of $\hat \psi$ is finite only 
for superpositions of states with different numbers of bosons.
The present work justifies taking 
such a superposition as a pure state of a subsystem.
More concretely, 
the expectation value of $\hat \psi$ is finite for the CSIB, 
$\langle \alpha, {\rm G}| \hat \psi |\alpha, {\rm G} \rangle ={\cal O}(1)$,
whereas it vanishes for the NSIB,
$\langle N, {\rm G}| \hat \psi |N, {\rm G} \rangle=0$ \cite{SMprl2000,SI}.
Although both the NSIB and CSIB have the off-diagonal 
long-range order \cite{SMprl2000,SI}, the gauge symmetry is broken, 
in the sense that $\langle \hat \psi \rangle ={\cal O}(1)$,
only for the CSIB \cite{ferro}. 
Since the state vectors of the NSIB and CSIB are 
quite different, they have different properties.
The most striking difference is that 
the NSIB decoheres much faster than the CSIB
when bosons have a finite probability of 
leakage into a huge 
environment, whose boson density is zero \cite{SMprl2000}.
Moreover, it was shown that 
the CSIB has the `cluster property,'
which ensures that fluctuations of any 
intensive variables are negligible, 
in consistency with thermodynamics \cite{SMcluster}.
Although the CSIB may look against the CSSR, 
the present work justifies taking it as a pure state of a subsystem, 
which, unlike the NSIB, is robust, symmetry breaking, and consistent with 
thermodynamics.
In particular, generalizing Eq.\ (\ref{CSIBkl}), 
one can take CSIB's for all subsystems.
In this case, the joint subsystem S is in a pure state 
in which each subsystem is in a CSIB.
Namely, the state vector of S behaves locally as a CSIB, 
which is robust, symmetry breaking, and consistent with 
thermodynamics.
Moreover, unlike Eq.\ (\ref{ent}), 
one can controle states of individual susbsystems
independently by local operations.
These justify macroscopic theories such as 
the Ginzburg-Landau theory and 
the Gross-Pitaevskii theory, which
assume, sometimes implicitly, that 
the order parameters can be defined locally,
and that 
their fluctuations are negligible, 
and that the state vectors are robust against weak perturbation 
from environments,
and that local operations do not cause global changes.
We consider that such a locally-CSIB state should be (a good approximation to)
a quantum state of real physical systems at low temperature, 
except for extreme cases
such as bosons in a perfectly-closed box at a ultra-low temperature. 
Finally, we remark that 
we have not assumed that the particles in the environment 
are in (or, not in) a condensed phase.


We thank H.\ Tasaki and M.\ Ueda for discussions.


\begin{references}



\bibitem[*]{shmz} Electronic address: shmz@ASone.c.u-tokyo.ac.jp


\bibitem{haag}
See, e.g., 
R.\ Haag, {\it Local Quantum Physics} (Springer, Berlin, 1992).

\bibitem{super}
See, e.g.,
M.\ Tinkham, 
{\it Superconductivity} 2nd ed.\ (McGraw-Hill, New York, 1996).

\bibitem{He4bubble}
E.\ G.\ Syskakis, F.\ Pobell and H. Ullmaier, 
Phys. Rev. Lett. {\bf 55} (1985) 2964.

\bibitem{atom}
M.\ H.\ Anderson {\it et al.},
Science {\bf 269} (1995) 198;
C.\ C.\ Bradley {\it et al.},
Phys. Rev. Lett. {\bf 75} (1995) 1687;
K.\ B.\ Davis {\it et al.},
Phys.\ Rev.\ Lett.\ {\bf 75} (1995) 3969.

\bibitem{andrews}
M.\ R.\ Andrews {\it et al.}, 
Science {\bf 275} (1997) 637.

\bibitem{java}
J.\ Javanainen and S.\ M.\ Yoo,  
Phys. Rev. Lett. {\bf 76} (1996) 161.

\bibitem{barnett}
S.\ M.\ Barnett, K.\ Burnett and J.\ A.\ Vaccaro, 
J.\ Res.\ Natl.\ Inst.\ Stand.\ Technol.\ {\bf 101}, 593 (1996).

\bibitem{SMprl2000}
A. Shimizu and T. Miyadera, 
Phys. Rev. Lett. {\bf 85}, 688 (2000).

\bibitem{range}
We assume that the product
$\prod_k C^{(k)}_{N_k n_k}$ is finite
only when $N_{\rm S} \ll N_{\rm tot}$.
Since E is assumed to be much larger than S, 
states with low energies would satisfy this condition.

\bibitem{vs}
A vector state is a state represented by a vector (more precisely, ray)
in a Hilbert space, which is not necessarily an irreducible 
representation: 
Since the Hilbert space consists of many charge sectors,
it becomes an irreducilbe representation of the algebra of
observables only when the observables include 
non-gauge invariant ones (such as a field operator),
which connect different sectors \cite{haag}.

\bibitem{precise}
Let us denote a quantum state symbolically by 
$\omega$ and the expectation value of an 
observable $A$ by $\omega(A)$.
A state $\omega$ is called {\it mixed}
iff there exist
states $\omega_1$ and $\omega_2$ ($\neq \omega_1$), 
and a positive number $\lambda$ ($0<\lambda<1$), 
such that 
$ 
\omega(A) = \lambda \omega_1(A) + (1-\lambda) \omega_2(A)
$ 
for any {\em local} observable $A$. 
Otherwise, $\omega$ is called a pure state.
See, e.g., Ref.\ \cite{haag}.

\bibitem{ent}
Here, the entanglement is defined by the increase of 
the von Neumann entropy for the reduced density operator
which is obtained by tracing over a part of subsytems.

\bibitem{peres}
See, e.g., A.\ Peres, 
Phys.\ Scripta T76, {\bf 52} (1998) (quant-ph/9707026).

\bibitem{perturbed}
On the other hand, the environment E might be 
perturbed by the control of a subsystem because 
$| \Phi \rangle_{\rm tot}$ is not a 
simple product of $\hat \rho_{\rm S}$ and a state vector of E.
However, we do not measure effects of such perturbations on E
because we assume that we do not measure anything of E.

\bibitem{SI}
A.\ Shimizu and J.\ Inoue, 
Phys.\ Rev.\ A {\bf 60}, 3204 (1999).

\bibitem{complicated}
Although this relation is the same as the relation between the 
corresponding states of free bosons,
$| \alpha, {\rm G} \rangle$ is a complicated wave function
because $|N, {\rm G} \rangle$ is complicated \cite{SMprl2000}.


\bibitem{ex-ent}
For example, the state (\ref{ent}) appears for two condensates,
each of which initially has a definite number of bosons, 
when an interference pattern is developed and observed.
See, e.g., J.\ I. Cirac et al., 
Phys.\ Rev.\ {\bf A54}, R3714 (1996).


\bibitem{hc}
If necessary, one can make
$\hat A_{\rm S}$ self-adjoint by adding its conjugate operator.


\bibitem{ferro}
This is a generic property of quantum phase transitions.
For example, if we treat a ferromagnetic Ising spin system as a quantum 
spin system, 
then two ferromagnetic states, 
$|\uparrow \uparrow \cdots \rangle$ and 
$|\downarrow \downarrow \cdots \rangle$, 
correspond to CSIB's, 
whereas their superpositions
$(|\uparrow \uparrow \cdots \rangle \pm 
|\downarrow \downarrow \cdots \rangle)/\sqrt{2}$
correspond to NSIB's:
Although all these states have a long-range order,
the up-down symmetry is broken only 
for the former states, in the sense that $\langle M_z \rangle \neq 0$.

\bibitem{SMcluster}
A. Shimizu and T. Miyadera, 
cond-mat/0009258.


\end{references}
\end{document}